\pdfoutput=1
\documentclass[11pt]{article}

\usepackage[]{ACL2023}
\usepackage{times}
\usepackage{latexsym}
\usepackage{graphics}
\usepackage{graphicx}
\usepackage{color}
\usepackage{xcolor}
\usepackage{colortbl}
\usepackage{caption}
\usepackage{amsfonts}
\usepackage{amssymb}
\usepackage{amsmath}
\usepackage{nicefrac}
\usepackage{bbm}
\usepackage{booktabs}
\usepackage{microtype}
\usepackage{multirow}
\usepackage{multicol}
\usepackage{tabu}
\usepackage{amsthm}
\usepackage{footmisc}
\usepackage{soul}
\usepackage{arydshln}
\usepackage{wasysym}
\usepackage{array}
\usepackage{makecell}
\usepackage{seqsplit}
\usepackage{algpseudocode}
\usepackage{algorithm}
\usepackage{hyperref}
\usepackage{xspace}
\usepackage{url}
\usepackage{utfsym}
\usepackage{dirtytalk}

\usepackage{float}
\usepackage{enumitem}
\usepackage{tikz}
\usepackage{tcolorbox}
\usepackage{siunitx}
\usepackage{pifont}% http://ctan.org/pkg/pifont
\usepackage{longtable}
\usepackage{supertabular}
\usepackage[fixed]{fontawesome5}

\newtcbox{\hlprimarytab}{on line, rounded corners, box align=base, colback=c3!10,colframe=white,size=fbox,arc=3pt, before upper=\strut, top=-2pt, bottom=-4pt, left=-2pt, right=-2pt, boxrule=0pt}
\newtcbox{\hlsecondarytab}{on line, box align=base, colback=red!10,colframe=white,size=fbox,arc=3pt, before upper=\strut, top=-2pt, bottom=-4pt, left=-2pt, right=-2pt, boxrule=0pt}
\newcommand{\daugshifted}{\raisebox{0.5\depth}{$\uparrow$}}
\newcommand{\uaugshifted}{\raisebox{0.5\depth}{$\downarrow$}}
\newcommand{\daulg}[1]{{\hlsecondarytab{\daugshifted{#1}}}}

\definecolor{my_bright_green}{RGB}{0,255,0}

\newtcbox{\hlgreentab}{on line, box align=base, colback=my_bright_green!20,colframe=white,size=fbox,arc=3pt, before upper=\strut, top=-2pt, bottom=-4pt, left=-2pt, right=-2pt, boxrule=0pt}

\newcommand{\daulggreen}[1]{{\hlgreentab{\uaugshifted{#1}}}}

\definecolor{my_green}{RGB}{51,102,0}
\definecolor{my_red}{RGB}{204, 0, 0}
\usepackage[T1]{fontenc}
\usepackage[utf8]{inputenc}

\usepackage{microtype}

\usepackage{inconsolata}

\usepackage{todonotes}
\title{
\protect 
\includegraphics[scale=0.06]{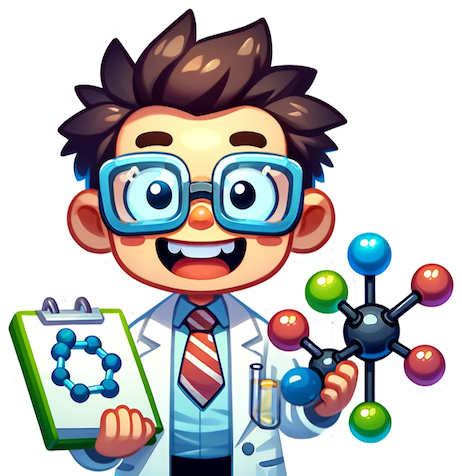} DRAK: Unlocking Molecular Insights with Domain-Specific Retrieval-Augmented Knowledge in LLMs
}
\author{
  Jinzhe Liu$^{\heartsuit \diamondsuit \spadesuit }$, 
  Xiangsheng Huang$^{\diamondsuit}$\footnotemark[1]\thanks{\quad Corresponding Author.}, 
  Zhuo Chen$^{\clubsuit}$,
  Yin Fang$^{\clubsuit}$\\
  $^\heartsuit$Institute of Computing Technology, Chinese Academy of Sciences\\
  $^\diamondsuit$Xiongan Institute of Innovation, Chinese Academy of Sciences\\
  $^\spadesuit$University of Chinese Academy of Science\\
  $^\clubsuit$College of Computer Science and Technology, Zhejiang University\\
 \texttt{ 
    \{liujinzhe23\}@mails.ucas.ac.cn
  }\\
}

\begin{document}
\maketitle
\begin{abstract}

Large Language Models (LLMs) encounter challenges with the unique syntax of specific domains, such as biomolecules.
Existing fine-tuning or modality alignment techniques struggle to bridge the domain knowledge gap and understand complex molecular data, limiting LLMs' progress in specialized fields.
To overcome these limitations, we propose an expandable and adaptable non-parametric knowledge injection framework named \textit{\textbf{D}omain-specific \textbf{R}etrieval-\textbf{A}ugmented \textbf{K}nowledge} (\textbf{DRAK}), aimed at enhancing reasoning capabilities in specific domains.
% Integrating a domain knowledge base with the LLMs aids in a more nuanced and context-aware analysis of molecular information. 
Utilizing knowledge-aware prompts and gold label-induced reasoning, DRAK has developed profound expertise in the molecular domain and the capability to handle a broad spectrum of analysis tasks.
% We constructed and evaluated the beneficial effects of two different incentive strategies on the performance of LLMs, achieving state-of-the-art results on six molecular tasks within the Mol-Instructions dataset. 
We evaluated two distinct forms of DRAK variants, proving that DRAK exceeds previous benchmarks on six molecular tasks within the Mol-Instructions dataset. Extensive experiments have underscored DRAK's formidable performance and its potential to unlock molecular insights, offering a unified paradigm for LLMs to tackle knowledge-intensive tasks in specific domains. Our code will be available soon.
% This highlights the advantages and scalability of our strategy, providing a viable approach for applying LLMs in the context of specific domains.

% \footnote{Code will be available at \url{https://github.com/zjunlp/AutoAct}.}.
%Further analysis demonstrates the effectiveness of the \textit{division-of-labor} strategy, with the trajectory quality generated by {\ours} significantly outperforming that of other methods
\end{abstract}

\section{Introduction}
LLMs are powerful parametric tools, equipped with extensive worldly knowledge and simulation capabilities.
Scholars are actively exploring the practical application potential of LLMs in knowledge-intensive domains such as finance, law, and biomolecules for 
% real-world tasks~\cite{survey/1,survey/2,survey/3}. 
real-world tasks~\cite{survey/3,DBLP:journals/corr/abs-2310-06671,Mol-Instructions,DBLP:journals/corr/abs-2202-10587}. 
Specifically within the biomolecules domain, LLMs employ self-supervised learning on extensive unlabeled datasets to identify novel drug candidates and predict drug synthesis pathways~\cite{bioLLM/1,bioLLM/3,bioLLM/2}. 
Building on these strengths, our goal is to develop a biomolecule-focused AI, similar to ChatGPT, to revolutionize professionals' interaction with molecular data. 

\begin{figure}[t!]
    \centering
    \includegraphics[width=\columnwidth]{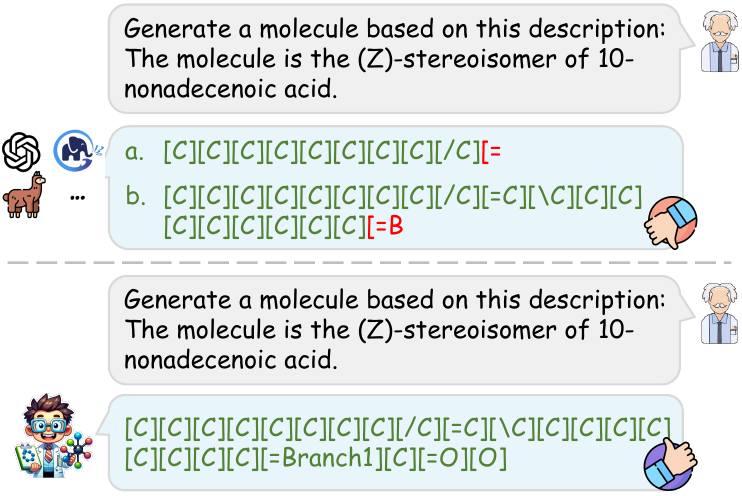}
    \caption{\textbf{DRAK:} Enabling Precise and Comprehensive Molecular Representations.}
    \label{fig:intropic}
\end{figure}
However, the reliability of LLMs decreases in tasks with long-tail knowledge within specific domains, due to a lack of domain knowledge and a propensity for ``\textit{hallucinations}''~\cite{hallucination/1, hallucination/2}, which can lead to severe consequences in critical applications.
% Within general domains, the constraints on responses are typically less stringent, permitting a range of correct answers and a certain level of inaccuracy.
% Conversely, w
% When dealing with knowledge within particular domains, LLMs are required to exhibit advanced cognitive and analytical skills, such as precisely identifying the structures and properties of molecules and proteins, as depicted in Figure \ref{fig:intropic}.
% Existing LLMs, trained predominantly on vast standard text corpora, are often limited by tokenizers optimized for English, struggling with non-standard formats like molecular structures due to inadequate domain knowledge. 
As shown in Figure \ref{fig:intropic}, when dealing with knowledge in specific domains, LLMs need to demonstrate advanced cognitive and analytical skills, such as accurately identifying the structures and properties of molecules and proteins. 
A significant challenge is that existing LLMs, primarily trained on extensive standard text corpora, are often constrained by tokenizers optimized for human language, making it difficult to handle non-standard formats like molecular structures due to insufficient domain knowledge.
Current methods, including fine-tuning and modality alignment, fall short of fundamentally resolving this challenge. 
To bridge this knowledge gap, integrating domain-specific knowledge is essential. 
Our method has improved the accuracy and comprehensiveness of LLM in predicting domain-specific knowledge, offering valuable insights to experts.

\begin{figure*}[t!]
    \centering
    \includegraphics[width=\textwidth]{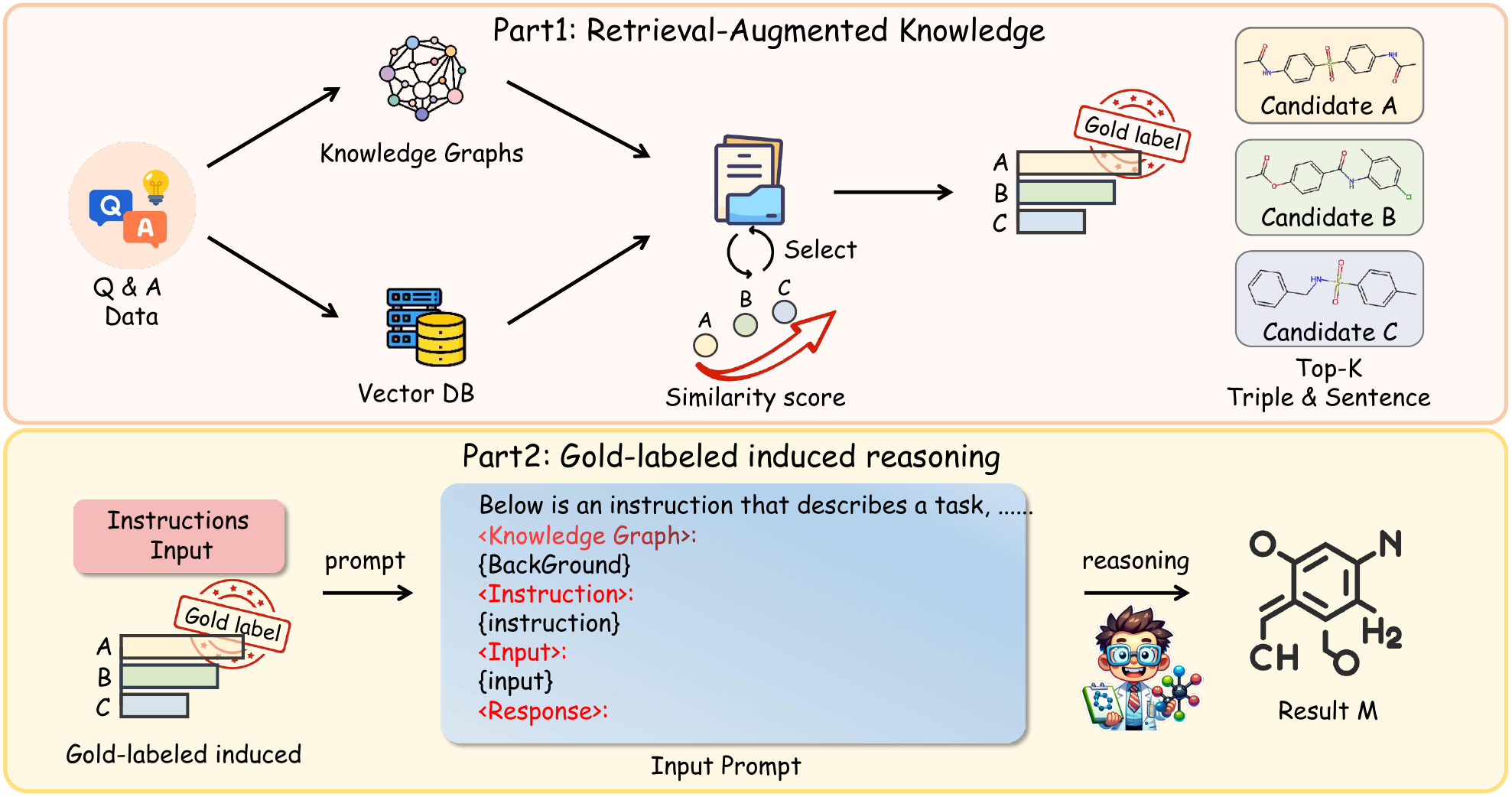}
    \caption{\textbf{The overview of our proposed framework DRAK.} \textbf{Part 1:} Top-K demonstration retrieval for enhanced molecular insight. \textbf{Part 2:} Gold-label induced reasoning.}
    \label{fig:method}
\end{figure*}

Knowledge injection has emerged as a crucial technique for empowering LLMs to produce dependable responses and grasp domain-specific knowledge~\cite{RAK/1, RAK/2}. 
Applying \textit{\textbf{D}omain-specific \textbf{R}etrieval-\textbf{A}ugmented \textbf{K}nowledge} (\textbf{DRAK}) to real-world scenarios poses main challenges: 1) \textit{improving LLMs' proficiency in leveraging background knowledge for content generation that meets specific criteria;} 2) \textit{comprehensively evaluating DRAK variants for optimal benefit.}
Essentially, these challenges revolve around the fundamental issue of LLMs' understanding of domain-specific knowledge. 
Taking the field of biomolecules as an example, we have designed a non-parametric, training-free domain knowledge injection framework as a general paradigm to address the inherent domain-specific QA challenges in practical LLM applications. 
Additionally, we analyze the advantages of injecting different forms of DRAK (\textbf{DRAK-S}, \textbf{DRAK-K}) and elucidate how to achieve optimal augmentation of domain knowledge by integrating domain knowledge into the model's reasoning using knowledge-aware prompts and golden label induction, thereby enhancing its effectiveness and accuracy in domain-specific applications and providing tangible support for experts.
% Essentially, these challenges focus on the fundamental issue of graduate students' insights into domain-specific knowledge.
% Bridging this knowledge gap requires integrating domain-specific knowledge into the models' reasoning, thereby boosting their effectiveness and precision in domain-specific applications and offering tangible support to experts.
% Taking the biomolecules field as an example, we designed a non-parametric, training-free domain knowledge injection framework as a universal paradigm to address the inherent domain-specific QA challenges in real-world LLM applications. 
% Furthermore, we dissected the advantages of injecting different forms of DRAK (\textbf{DRAK-S}, \textbf{DRAK-K}) and elucidated how to achieve optimal augmentation of domain-specific knowledge.

The main contributions are as follows:
\begin{itemize}
    \item  \textit{\textbf{DRAK Impact Analysis:}} Highlights the distinct impacts of various DRAK forms in improving LLMs' mastery of knowledge, particularly in understanding and responding accurately to domain-specific queries.
    \item \textit{\textbf{Knowledge Graphs Integration:}} Explores integrating DRAK with knowledge graphs (KGs) in the biomolecular domain, employing structured non-parametric knowledge injection to enhance LLMs' understanding and generative capabilities in specialized areas.
    \item \textit{\textbf{Universal Application Framework:}} Presents a scalable and adaptable framework for implementing domain-expert LLMs in areas characterized by distinct syntactic structures (such as molecules, materials, and chemistry), leading to significant advancements in inference capabilities.
\end{itemize}

\section{Methodology}
\subsection{Definition of Concepts and Problems}
\paragraph{Knowledge Graph Definition.}
Following \citet{DBLP:journals/corr/abs-2402-05391}, a KG is defined as \( \mathcal{G}=\{(e, r, e') | e, e' \in \mathcal{E}, r \in \mathcal{R}\} \) consists of entities \( \mathcal{E} \), relations \( \mathcal{R} \), and triples \( (e, r, e') \) representing factual relationships. It provides a structured background knowledge base for molecular tasks, enriching LLMs with domain-specific information.
\paragraph{Problem Definition.}
We propose DRAK as \(\mathcal{DRAK}=\mathcal{Q} \times \mathcal{K} \rightarrow \mathcal{A}\), integrated into the objective \(\mathcal{F}\), defined as \(\mathcal{F}=\mathcal{T} \times \mathcal{L} \times \mathcal{D} \rightarrow \mathcal{O}\), aiming to enhance LLMs' understanding and accuracy in molecular tasks. 
Here, \(\mathcal{Q}\) represents the set of queries, \(\mathcal{G}\) denotes the background knowledge, \(\mathcal{A}\) is the set of answers, \(\mathcal{T}\) is the set of molecular tasks, \(\mathcal{L}\) indicates the collection of LLMs, \(\mathcal{D}\) is the set of \(\mathcal{DRAK}\), and \(\mathcal{O}\) is the output set for tasks. 
This unified framework seeks to utilize DRAK to deepen the comprehension and improve the predictive accuracy of Large Language Models in tasks related to the molecular domain, successfully addressing the shortcomings of conventional approaches.

\subsection{Prompt Construction}
\paragraph{Instructions \(\mathcal{I}\) and Input \(\mathcal{M}\).}
For AI-aided generation in biomolecules, we leverage PubMed~\cite{pubmed} scientific abstracts to create open-ended question prompts \(\mathcal{I}\).
Consistent with Mol-Instruction~\cite{Mol-Instructions}, gpt-3.5-turbo is adopted to enrich the form of Instruction. 
For input \(M \), we integrate external knowledge and \(\mathcal{M}\) into prompt through templates.
\paragraph{Demonstration \(\mathcal{D}\).}
Initially, we utilize DRAK to acquire a demonstration set \(D\) tailored for each test case. 
To explore the sensitivity of LLMs towards molecular strings, we design two distinct variants of in context learning (ICL) demonstrations. 
The aim is to determine the LLM's ability to accurately understand molecular structures. 
Each demonstration \( (x_i, y_i) \) is augmented with gold label data from the training set, abstracted into a new set \( (x_i, y_i, r_i) \), designated as \(\mathcal{D}_i\), where \(x_i\) and \(y_i\) represent the input and output, respectively, and \(r_i\) embodies the inferred association between them.

\subsection{Retrieval-Augmented Knowledge}
Retrieval-augmented techniques, leveraging textual similarity embeddings close to test samples, enhance consistency and robustness across diverse scenarios~\cite{liu2021makes}.
However, LLMs lacks sufficient accuracy in predicting entities with subtle structural differences, such as chemical molecular formulas represented by SMILES~\cite{SMILES} or SELFIES~\cite{SELFIES}.
To address this challenge, we consider two core aspects:
1) The format of prompts inspired by chain-of-thought~\cite{CoT}. We incorporate textual reasoning chains to introduce external knowledge for generating reliable and specific reasoning processes.
2) Inspired by triple structures in KG~\cite{zhang2023knowledgeable}, we enhance prompts with structured features.
We examine the positive influences of utilizing \textit{Sentence-centric Knowledge Injection} (\textbf{DRAK-S}) and \textit{KG-driven Knowledge Injection} (\textbf{DRAK-K}) on LLMs, as detailed in \S 2.3.1 and \S 2.3.2, respectively.

\subsubsection{Sentence-Centric Knowledge Injection}
Acknowledging the pivotal role that molecular entity information plays in sentence comprehension, we have formulated a \textbf{DRAK-S} strategy for the integration of entity-level weakly labeled information to facilitate context reconstruction.
Utilizing the task of text-described molecular generation as an illustrative example, envisage the scenario: \textit{The molecule is a natural product discovered in Cytophaga, with data accessible.} This scenario aligns with the molecule's SELFIES structural representation: ``\texttt{\seqsplit{[C][C][Branch1][C][C][Branch1][Ring1][C][S][S]}}''.
This is transmuted into a natural language assertion: \textit{The molecular structure corresponding to the aforementioned description is:} ``\texttt{\seqsplit{[C][C][Branch1][C][C][Branch1][Ring1][C][S][S]}}''.
The objective is to assess the adeptness of LLMs in proficiently comprehending and parsing a confluence of natural textual content and molecular representations.
Specifically, by introducing quotation-marked SELFIES~\cite{SELFIES} strings, we provide LLM with intuitive cues for molecular structure. 
This method preserves both the sentence's semantic integrity and the accuracy of information related to entity pairs.
Finally, the vector similarity of sentence embedding is calculated to achieve accurate knowledge recall.

We set the similarity score threshold filtering (\(\mathit{SF}\)) method to recall the gold label data closest to the query, and the formula is as follows:
% \begin{equation}
% \begin{aligned}
% \mathit{SF} &= \underset{i=1}{\mathit{argmax}} \left\{ \mathit{sim}(x, x_i) \mid \mathit{sim}(x, x_i) \geq \theta \right\}
% \end{aligned}
% \end{equation}
\begin{equation}
    \mathit{SF} = \underset{i=1}{\mathit{argmax}} \left\{ \mathit{sim}(x, x_i) \right\}
\end{equation}
where \( \mathit{sim}(x, x_i) \geq \theta \). The formula identifies the instance \( i \) maximizing the similarity $sim(x,x_i)$, provided that this similarity exceeds a predefined threshold \( \theta \). 
This ensures that only the data closely aligned with the query, in terms of content and context, is considered for recall, enhancing the precision of our retrieval process.

\subsubsection{KG-driven Knowledge Injection}
Reasoning within existing knowledge systems is adeptly captured through induction and deduction, mirroring the human approach of employing mental maps for problem-solving~\cite{DBLP:conf/semweb/0007CGPYC21,DBLP:journals/corr/abs-2402-05391}. The triplet structure of knowledge bases (\textit{subject, relation, object}) methodically formalizes this reasoning process.
Unlike sentence-level knowledge representation, \textbf{DRAK-K} approach endeavors to furnish LLMs with data prompts that are more structured and meticulously refined.
Entity matching retrieval is performed through querying and task instructions, recalling the top-k most relevant KG, delineated as a list of triplets \( \mathcal{G}_k \).

For example, for the triplet (``\textit{The molecule is a natural product found in Cytophaga with data available.}'', ``\textit{description guided molecule design}'', ``\texttt{\seqsplit{[C][C][Branch1][C][C][Branch1][Ring1][C][S][S]}}''), the instruction \( \mathcal{I} \) ``\textit{Generate a molecule based on this description.}'' is mapped to ``\textit{description guided molecule design}''. 
The knowledge graph \( \mathcal{G} \) is transformed into structured background context for LLMs, clearly describing the task-relevant molecular structures and their application contexts.

\subsection{Language Model Reasoning}
We employ a vector database to construct expert knowledge priors, safeguarding domain-specific data's security and confidentiality. 
This endows the LLM with non-parametric memory capabilities, and it is easily portable and updatable.
\paragraph{Gold Label-Induced.} We explored a gold-label-induced (GLI) reasoning method aimed at providing LLMs with a carefully curated set of a few demonstrations by extracting the most similar examples from the training set for each test input based on data distribution~\cite{wan2023gpt}.

Specifically, for each test input \( x \), we define a similarity function \( \mathit{sim}(x, x_i) \), where \( x_i \) is an instance in the training set, and \( \mathit{sim} \) quantifies the similarity between two instances. 
We select \( k \) instances \( \{ x_{i1}, x_{i2}, ..., x_{ik} \} \) exhibiting the highest similarity scores and furnish their corresponding gold labels \( \{ y_{i1}, y_{i2}, ..., y_{ik} \} \) as essential background information to the LLMs.
This process can be represented by the following formula:
\begin{equation}
\begin{aligned}
\mathit{GLI}(x) = \underset{\{x_{i1}, ..., x_{iN}\} \subseteq X}{\mathit{argmax}} \sum_{j=1}^{N} \mathit{sim}(x, x_{ij})
\end{aligned}
\end{equation}
where \( X \) represents the set of all instances in the training set, and \( \mathit{GLI}(x) \) represents the set of \( k \) most similar instances retrieved for input \( x \).

This method enables Large Language Models (LLMs) to utilize exact, contextually pertinent gold label data prior to formulating responses, thereby guaranteeing both precision and contextual appropriateness. 
The GLI method stands out in knowledge-dense areas such as legal consulting and medical diagnosis by refining LLM responses through similarity matching against a constrained training dataset.
DRAK enhances LLMs by conducting top-k gold standard data similarity retrieval using directive and input pairs \( \langle inst, Inp\rangle \), aiding in producing accurate and contextually relevant answers while minimizing hallucinations.

% \input{tabs/main_results}

% \paragraph{Language Model Reasoning}
% DRAK provides a non-parametric knowledge injection method that bypasses training, using instruction and input pairs \( \langle inst, Inp\rangle \) for vector similarity searches to enrich LLMs with context for generating sequences \( y \).
% The process begins by creating task-specific prompts from two types of examples and then uses gold label data for retrieval, enabling the recall of top-K questions as core demonstrations.
% DRAK offers LLMs a flexible external source of knowledge, assisting them in generating accurate and contextually relevant answers while reducing the occurrence of hallucinations.

\begin{table*}[htbp] % This ensures the table spans across both columns
\centering
% \resizebox{\textwidth}{!}{%
\resizebox{\textwidth}{!}{%
\begin{tabular}{lccccccc}
\toprule
\rowcolor[gray]{0.9}
\textbf{MODEL} & \textbf{EXACT$\uparrow$} & \textbf{BLEU$\uparrow$} & \textbf{LEVENSHTEIN$\downarrow$} & \textbf{RDK FTS$\uparrow$} & \textbf{MACCS FTS$\uparrow$} & \textbf{MORGAN FTS$\uparrow$} & \textbf{VALIDITY$\uparrow$} \\
\midrule
\multicolumn{8}{c}{\textbf{Description-guided Molecule Design}} \\
\midrule[1.1pt]
\rowcolor[RGB]{234, 238, 234} \multicolumn{8}{l}{\textit{Specialist Models}} \\
TEXT+CHEM T5 & 0.097 & 0.508 & 41.819 & 0.352 & 0.474 & 0.353 & 0.721 \\
MOLT5        & \textbf{0.112} & \textbf{0.546} & 38.276 & 0.400 & 0.538 & 0.295 & 0.773 \\
\midrule
\rowcolor[RGB]{234, 238, 234} \multicolumn{8}{l}{\textit{LLM Based Generalist Models}} \\
\midrule
ALPACA       & 0.000 & 0.004 & 51.088 & 0.006 & 0.029 & 0.000 & 0.022 \\
BAIZE        & 0.000 & 0.006 & 53.796 & 0.000 & 0.000 & 0.000 & 0.017 \\
CHATGLM      & 0.000 & 0.004 & 53.157 & 0.005 & 0.000 & 0.000 & 0.046 \\
LLAMA        & 0.000 & 0.003 & 59.864 & 0.000 & 0.000 & 0.000 & 0.026 \\
VICUNA       & 0.000 & 0.006 & 60.356 & 0.006 & 0.001 & 0.000 & 0.011 \\
GALACTICA    & 0.000 & 0.192 & 44.152 & 0.135 & 0.248 & 0.088 & 0.992 \\
MOL-INSTRUCTIONS        & 0.002 & 0.345 & 41.367 & 0.231 & 0.412 & 0.147 & 1.000 \\
\textbf{DRAk-S (ours)}             & 0.049 & 0.392 & 35.726 & 0.341 & 0.497 & 0.235 & 1.000 \\
\textbf{DRAk-K (ours)}               & \textbf{0.104} & \textbf{0.515} & \textbf{32.641} & \textbf{0.455} & \textbf{0.600} & \textbf{0.326} & \textbf{1.000} \\
\midrule
\multicolumn{8}{c}{\textbf{Reagent Prediction}} \\
\midrule[1.1pt]
\rowcolor[RGB]{234, 238, 234} \multicolumn{8}{l}{\textit{Specialist Models}} \\
TEXT+CHEM T5 & 0.000 & 0.225 & 49.323 & 0.039 & 0.186 & 0.052 & 0.313 \\ 
\midrule
\rowcolor[RGB]{234, 238, 234} \multicolumn{8}{l}{\textit{LLM Based Generalist Models}} \\
\midrule
ALPACA       & 0.000 & 0.026 & 29.037 & 0.029 & 0.016 & 0.001 & 0.186 \\
BAIZE        & 0.000 & 0.051 & 30.628 & 0.022 & 0.018 & 0.004 & 0.099 \\
CHATGLM      & 0.019 & 0.019 & 29.169 & 0.017 & 0.006 & 0.002 & 0.074 \\
LLAMA        & 0.000 & 0.013 & 28.040 & 0.037 & 0.001 & 0.001 & 0.001 \\
VICUNA       & 0.000 & 0.003 & 27.948 & 0.038 & 0.002 & 0.001 & 0.007 \\
GALACTICA    & 0.000 & 0.141 & 30.760 & 0.036 & 0.127 & 0.051 & 0.995 \\
MOL-INSTRUCTIONS            & 0.044 & 0.224 & 23.167 & 0.237 & \textbf{0.364} & \textbf{0.213} & 1.000 \\
\textbf{DRAk-S (ours)}         & 0.023 & 0.450 & 25.223 & 0.187 & 0.245 & 0.187 & 1.000 \\
\textbf{DRAk-K (ours)}          & \textbf{0.049} & \textbf{0.487} & \textbf{22.87} & \textbf{0.238} & 0.331 & 0.207 & \textbf{1.000} \\
\midrule
\multicolumn{8}{c}{\textbf{Forward Reaction Prediction}} \\
\midrule[1.1pt]
\rowcolor[RGB]{234, 238, 234} \multicolumn{8}{l}{\textit{Specialist Models}} \\
TEXT+CHEM T5 & 0.239 & \textbf{0.782} & 20.413 & \textbf{0.705} & \textbf{0.789} & \textbf{0.652} & 0.762 \\
\midrule
\rowcolor[RGB]{234, 238, 234} \multicolumn{8}{l}{\textit{LLM Based Generalist Models}} \\
\midrule
ALPACA       & 0.000 & 0.065 & 41.989 & 0.004 & 0.024 & 0.008 & 0.138 \\
BAIZE        & 0.000 & 0.044 & 41.500 & 0.004 & 0.025 & 0.009 & 0.097 \\
CHATGLM      & 0.000 & 0.183 & 40.008 & 0.050 & 0.044 & 0.108 & 0.039 \\
LLAMA        & 0.000 & 0.020 & 42.002 & 0.001 & 0.002 & 0.001 & 0.059 \\
VICUNA       & 0.000 & 0.057 & 41.690 & 0.016 & 0.006 & 0.001 & 0.059 \\
GALACTICA    & 0.000 & 0.468 & 35.021 & 0.156 & 0.257 & 0.097 & 0.946 \\
MOL-INSTRUCTIONS        & 0.045 & 0.654 & 27.262 & 0.313 & 0.509 & 0.262 & 1.000 \\
\textbf{DRAk-S (ours)}     & 0.038 & 0.739 & 26.090 & 0.418 & 0.546 & 0.325 & 1.000 \\
\textbf{DRAk-K (ours)}      & \textbf{0.254} & \textbf{0.778} & \textbf{18.649} & \textbf{0.602} & \textbf{0.741} & \textbf{0.546} & \textbf{1.000} \\
\midrule
\multicolumn{8}{c}{\textbf{Retrosynthesis}} \\
\midrule[1.1pt]
\rowcolor[RGB]{234, 238, 234} \multicolumn{8}{l}{\textit{Specialist Models}} \\
TEXT+CHEM T5 & 0.141 & 0.765 & 24.043 & \textbf{0.685} & \textbf{0.765} & \textbf{0.585} & 0.698 \\
\midrule
\rowcolor[RGB]{234, 238, 234} \multicolumn{8}{l}{\textit{LLM Based Generalist Models}} \\
\midrule
ALPACA       & 0.000 & 0.065 & 41.989 & 0.004 & 0.024 & 0.008 & 0.138 \\
BAIZE        & 0.000 & 0.044 & 41.500 & 0.004 & 0.025 & 0.009 & 0.097 \\
CHATGLM      & 0.000 & 0.183 & 40.008 & 0.050 & 0.044 & 0.108 & 0.039 \\
LLAMA        & 0.000 & 0.020 & 42.002 & 0.001 & 0.002 & 0.001 & 0.059 \\
VICUNA       & 0.000 & 0.057 & 41.690 & 0.016 & 0.006 & 0.001 & 0.059 \\
GALACTICA    & 0.000 & 0.452 & 34.940 & 0.167 & 0.274 & 0.134 & 0.986 \\
MOL-INSTRUCTIONS        & 0.009 & 0.705 & 31.227 & 0.283 & 0.364 & 0.213 & 1.000 \\
\textbf{DRAk-S (ours)}     & \textbf{0.400} & 0.760 & 31.118 & 0.339 & 0.507 & 0.265 & 1.000 \\
\textbf{DRAk-K (ours)}      & 0.319 & \textbf{0.793} & \textbf{20.779} & \textbf{0.625} & \textbf{0.758} & \textbf{0.565} & \textbf{1.000} \\
\bottomrule
\end{tabular}
}
\caption{Results of molecular generation tasks. These tasks encompass description-guided molecule design, reagent prediction, forward reaction prediction, and retrosynthesis.}
\label{tab:model_comparison}
\end{table*}

\section{Experiment}
\subsection{Experimental Setup}
\paragraph{Datasets and Metrics.}
We leverage the Mol-Instructions~\cite{Mol-Instructions} training set as gold-label data, evaluating model performance across tasks with 1k test samples from its test set.

To assess molecular understanding, we employ BLEU~\cite{BLEU}, ROUGE~\cite{ROUGE}, and METEOR~\cite{METEOR} metrics for evaluating output quality against reference answers. 
In molecule generation, validity is checked with RDKit~\cite{RDKit}, followed by exact match comparisons. Given the variability of molecular structures, we also measure molecular similarity using RDKit/MACCS/Morgan fingerprints~\cite{tanimoto,MACCS,Morgan}, alongside Levenshtein~\cite{Levenshtein} and BLEU scores. 
For molecular property prediction, MAE (mean absolute error) quantifies prediction accuracy of continuous values.

\begin{figure*}[t!]
    \centering
    \resizebox{1.\textwidth}{!}{
    \includegraphics{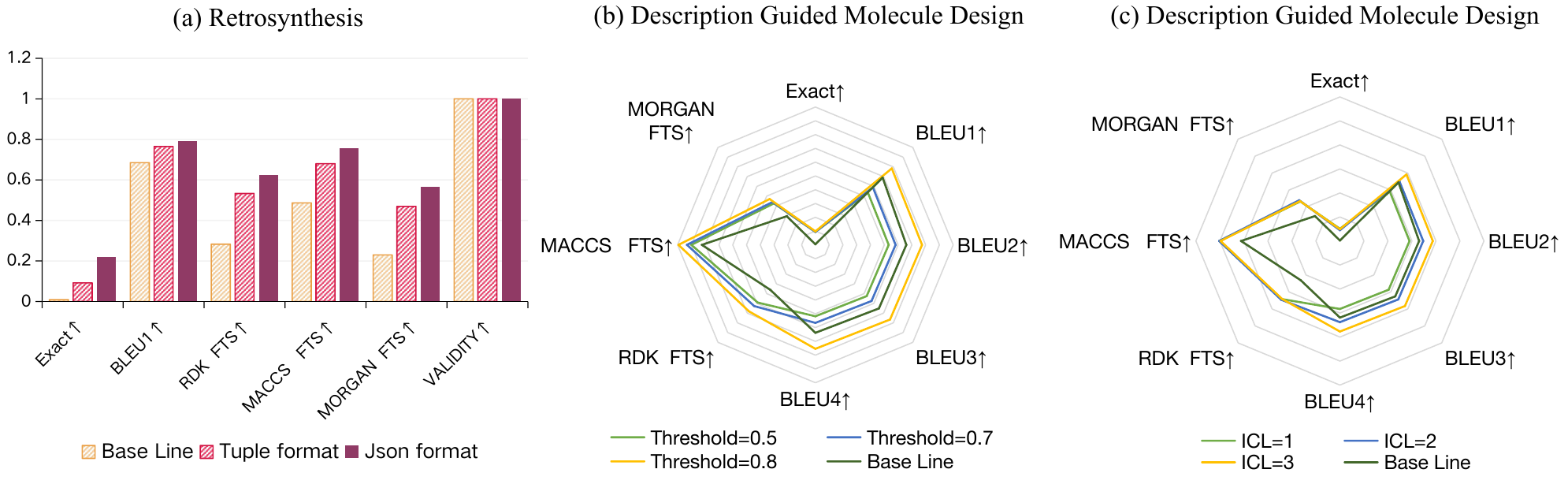}}
    \caption{Ablation Study of DRAK Across Different Knowledge Representation Formats}
    \label{fig:ablations}
\end{figure*}

% \vspace{5mm} 
\begin{figure}[t!]
    \centering
    \includegraphics[width=\columnwidth]{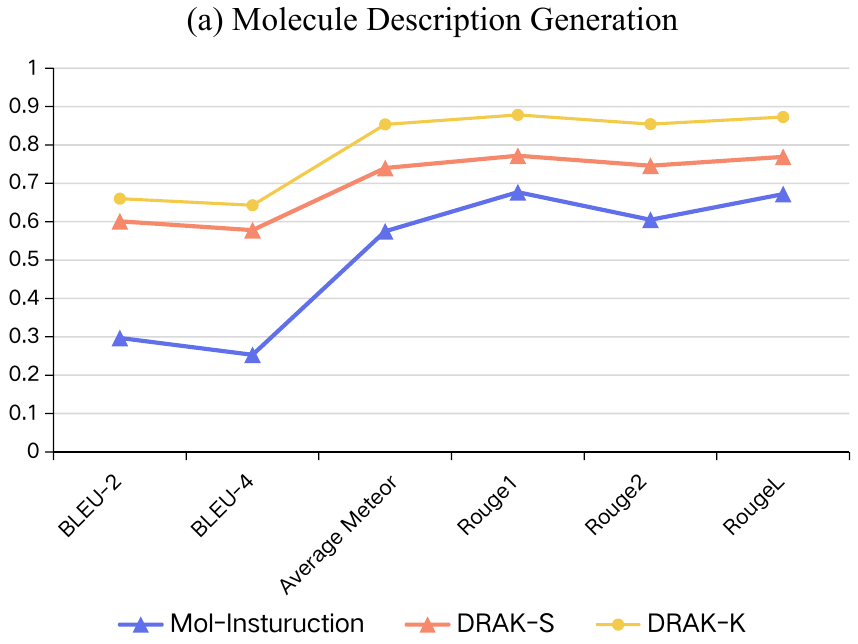}
    \caption{Results of molecular-description generation tasks. From top to bottom: DRAK-K, DRAK-S, Mol-Instructions.}
    \label{fig:moleculardescriptiongen}
\end{figure}

\paragraph{Implementation Details.}
The DRAK injection strategy is applied to a Mol-Instructions~\cite{Mol-Instructions} fine-tuned LLaMA-7B model, operating efficiently on a single NVIDIA RTX 3090 GPU, demonstrating its practicality for resource-constrained environments.
Leveraging the LangChain framework, we have refined our retrieval strategy with a threshold-based filtering mechanism, calibrated at a threshold of 0.8, to meticulously extract the top-k most relevant examples housed within our ChromaDB vector database.
Regarding the realization of KG, we use networkx framework to build and manage complex knowledge structure, so that our model can effectively use structured knowledge to enhance reasoning ability.
% Please refer to the Appendix for additional details.

\subsection{Overall Results}
We conducted a comprehensive comparison across six molecular tasks between Alpaca-LoRA~\cite{alpaca-lora}, Baize-7B~\cite{Baize}, ChatGLM-6B~\cite{Glm}, LLama-7B~\cite{llama}, Vicuna~\cite{vicuna}, Mol-Instructions~\cite{Mol-Instructions}, and two variants of DRAK proposed by us.

DRAK-S utilizes sentence-level prompts, while DRAK-K uses a KG for structured prompting.
Table \ref{tab:model_comparison} presents a comparison of DRAK-S and DRAK-K against benchmarks based on LLMs and expert models.
1) The DRAK-S and DRAK-K models consistently outperform various LLM-based benchmarks on metrics such as BLEU scores, molecular similarity, and exact match. 
DRAK-K demonstrated superior exact match scores compared to Mol-Instructions~\cite{Mol-Instructions} across four molecular generation tasks, achieving improvements of \daulg{0.102}, \daulg{0.005}, \daulg{0.209}, and \daulg{0.310}, respectively.
2) Compared to specialized small models that compromise on generalizability, DRAK, as a general model, exhibits competitive molecular generation capabilities and even surpasses expert models in the Reagent Prediction task comprehensively.
3) The GLI inference strategy improves the model's understanding in data-limited domains by enriching the context.

\begin{table}[!t]
\vskip -0.2in % 调整表格与上文的距离
\centering
\setlength{\tabcolsep}{9pt} % 将列间距减小到5pt
\renewcommand{\arraystretch}{1} % 减小行高
\begin{minipage}{\columnwidth} % 使用整个列宽
	\centering
    \vskip 0.1in
	\resizebox{0.7\textwidth}{!}{
        \tiny
		\begin{tabular}{lccccccc}
% \toprule
% \midrule
\hline
\textbf{Model} & \textbf{MAE} $\downarrow$ \\
% \midrule 
\hline
\rowcolor[RGB]{234, 238, 234}
\multicolumn{2}{l}{\textit{Property Prediction}} \\
Alpaca & 322.109\\
Baize & 261.343\\
ChatGLM & - \\
LLama & 5.553\\
Vicuna & 860.051 \\
Galactica & 0.568 \\
Mol-Instructions & 0.013 \\
% \midrule
\hline
DRAK-S (Ours) & 0.019 \\
DRAK-K (Ours) & \textbf{0.00096} \\
\hline
% \midrule
% \bottomrule
\end{tabular}}
\end{minipage}
\caption{Results of molecular property prediction tasks.}
\label{tab:property_pred}
\vspace{-0.35em} % 调整表格与下文的距离
\end{table}

Similarly, DRAK exhibits remarkable capabilities in molecular understanding and property prediction. 
The property prediction outcomes depicted in Table \ref{tab:property_pred} reveal DRAK-K's significant reduction in MAE scores by two orders of magnitude relative to baseline models, decreasing from 0.013 to 0.00096, a reduction of \daulggreen{0.012}.
In the molecular understanding task shown in Figure In the molecular understanding task illustrated in Figure \ref{fig:moleculardescriptiongen}, compared to the Mol-Instructions~\cite{Mol-Instructions}, DRAK-K achieved increases of \daulg{0.363}, \daulg{0.278}, and \daulg{0.200} in BLEU-2, Meteor, and ROUGE scores, respectively. 
The structured triplets of the KG help improve the model's understanding of complex molecular structures.

\begin{figure*}[t!]
    \centering
    \resizebox{1.\textwidth}{!}{
    \includegraphics{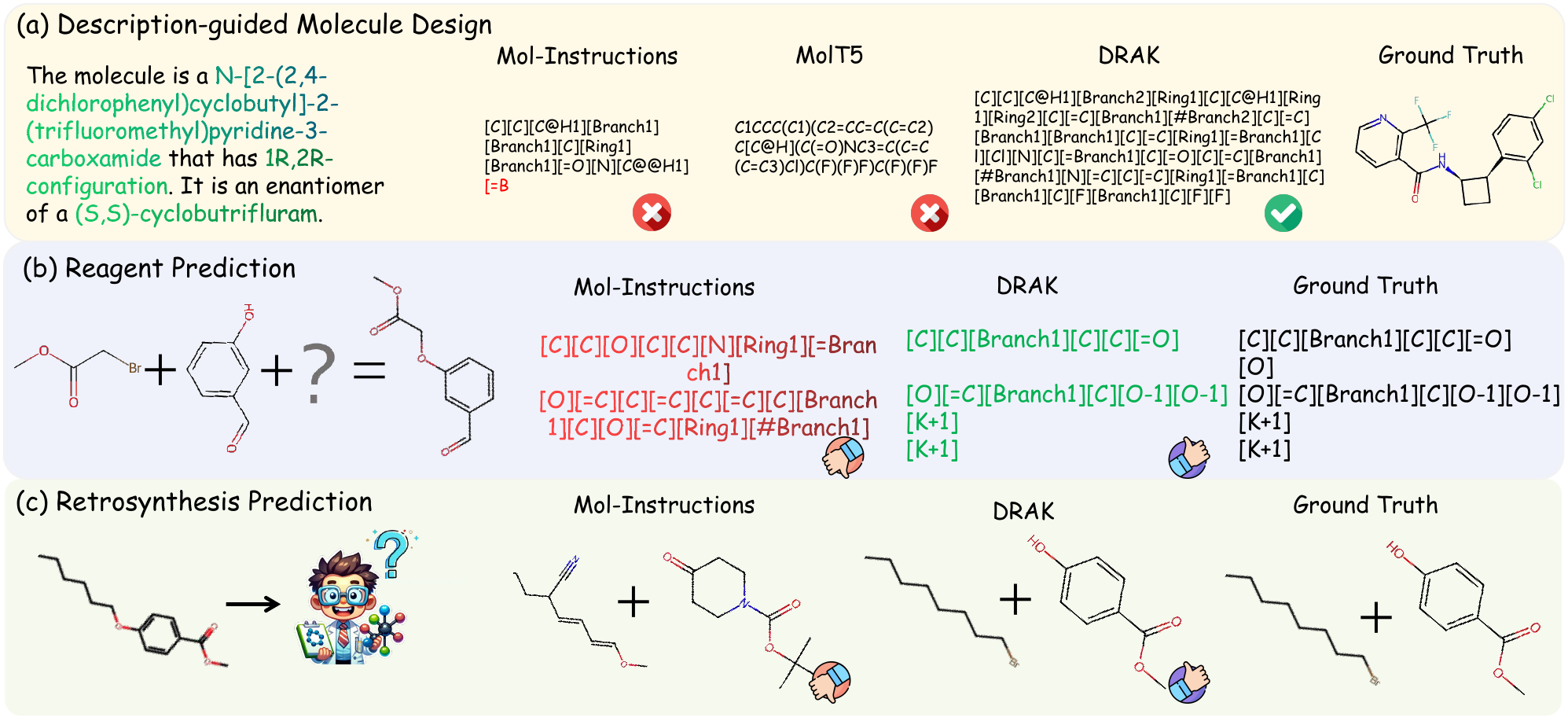}}
    \caption{Comparison of Description-guided Molecule Design and Chemical Reaction Task Results: (a) Demonstrates DRAK's precision in molecular structure responses. (b-c) Highlight DRAK's accurate generation of reaction-related compounds.
    }
    \label{fig:casestudy/1}
\end{figure*}

\begin{figure*}[t!]
    \centering
    \resizebox{1.\textwidth}{!}{
    \includegraphics{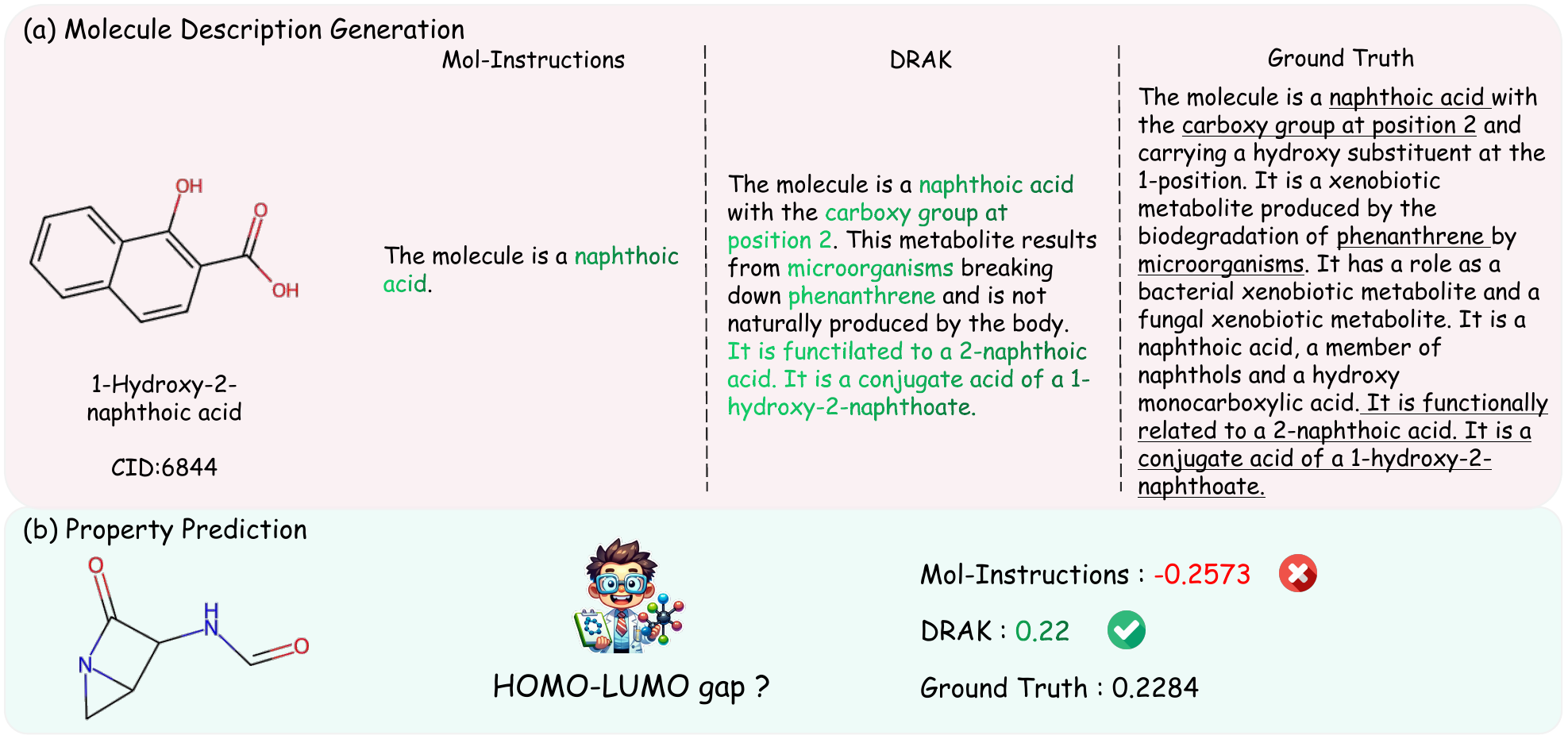}}
    \caption{Comparison of molecular description generation task and property prediction results. (a) reveals DRAK’s caption aligning closely with the ground truth. The colored text represents the exact match between the output and the ground truth. (b) DRAK's accuracy in property prediction.}
    \label{fig:casestudy/2}
\end{figure*}

In conclusion, DRAK combines retrieval-enhanced prompting and GLI strategy, excelling in molecular tasks and highlighting its precision in complex domains.

\subsection{Ablation Studies}
\paragraph{Q1: Is the result consistent with better templates?}
Beyond our minimalist KG triplet template, we also explored KG representations in JSON format.   
Figure \ref{fig:ablations} (a) depicts this trend, demonstrating that changes in KG representation have a minimal impact on performance, maintaining competitive results overall. 
Notably, the use of detailed templates does not always surpass minimalist templates, highlighting the sensitivity of LLMs to prompts.
\paragraph{Q2: Is the gold label improving the quality of RAK?}
Figure \ref{fig:ablations} (b) illustrates the retrieved labels at varying thresholds (0.5, 0.7, 0.8). As the threshold increases, vector retrieval brings back ICL examples with distributions that more closely resemble the test set. We observe a corresponding incremental enhancement in the model's response quality and evaluation metric scores with the rising threshold.
\paragraph{Q3: Is the result consistent with varying k?}
Contrary to conventional NLP tasks, Figure \ref{fig:ablations} (c) shows DRAK's limited improvement in molecular insight with a minimal number of examples ($k=1$). 
This may be due to the model's tendency to overly rely on a single instance based on ICL, failing to flexibly address the diversity of molecular structures and functions. 
As the number of instances increases, the model begins to grasp the common structural features of molecules under the same data distribution by analyzing a broader range of instances. 
This indicates that by incorporating more gold-label instances, DRAK can guide LLMs to discern molecular structures' common features, leading to high-quality responses.

\subsection{Case Studies}
Several examples of results across different tasks are illustrated in Figures \ref{fig:casestudy/1} and \ref{fig:casestudy/2}. 
Figure \ref{fig:casestudy/1} emphasizes the accuracy and comprehensiveness of DRAK in case studies involving description-guided molecule design, reagent prediction, and retrosynthesis prediction. The precise prediction of molecular reaction structures further demonstrates DRAK's effective capture of the intricate relationships between molecules and their functionalities, validating its profound understanding of complex molecular structures.
% Figure \ref{fig:casestudy/1} highlights DRAK's precision in generating compounds for chemical reaction predictions, underscoring its value in description-guided molecular design and retrosynthesis case studies. 
In Figure \ref{fig:casestudy/2}, part (a) showcases the molecule description generation task, where DRAK generates captions closely aligned with factual information, accurately identifying the molecular chemical structure(i.e., ``\textit{naphthoic acid with the carboxy group at position 2}''), functions (i.e., ``\textit{2-naphthoic acid, 1-hydroxy-2-naphthoate}''), and origins and production (i.e., ``\textit{microorganisms breaking down phenanthrene}'').
Part (b) pertains to the property prediction task, wherein DRAK relatively accurately predicts the molecule's \textit{HOMO-LUMO gap} energies based on its structure. In comparison, the results produced by Mol-Instructions demonstrate substantial discrepancies from the factual values.

% See Appendix for more details.
% Consistent with Mol-instruct, the molecular understanding task was evaluated by using BLEU(papi neni et al., 2002), ROUGE (Lin, 2004) and METEOR(baner JEE Lavie, 2005). For molecular generation, we first use rd kit(land rum et al., 2013) to verify whether the generated strings correspond to effective molecules, and then calculate their exact matches with reference solutions. To accommodate these complexities and provide a more comprehensive evaluation, we further adopted metrics measuring molecular similarity. These include similarity scores derived from RDKit/MACCS/Morgan fingerprints (Tanimoto, 1958; Schneider et al., 2015; Durant et al., 2002), as well as Levenstein (Yujian  Bo, 2007) and BLEU scores. For molecular property prediction tasks, we used MAE (Mean Absolute Error) to quantify the accuracy of model predictions for continuous values.}

\section{Related Work}
% In this section, we provide a literature review related to Molecule-related LLMs and Retrieval-Augmented Knowledge.
\subsection{Molecule-Related LLMs}
LLMs fine-tuning has demonstrated significant potential in transforming interactions with complex molecular data~\cite{survey/tuneing/1,survey/instruction,lora}.
MolT5~\cite{MolT5} illustrates LLMs' ability for cross-modal translation, molecular caption generation and text-driven design.
Mol-Instructions~\cite{Mol-Instructions} employs SELFIES descriptors to develop biomolecular-specific datasets, equipping general-domain LLMs with biomolecular expertise. 
InstructMol~\cite{InstructMol}, leveraging modality alignment strategies~\cite{Multi-modal/1, Multi-modal/2}, integrates 2D molecular graphs into LLMs with a pretrained encoder and cross-modal learning, bridging the gap between molecular structures and text.
This fine-tuning often prioritizes output format mimicry over complex structure comprehension, limiting generalization and elevating ``hallucination'' risks.  
Conversely, our DRAK approach seamlessly integrates knowledge, blending parametric and non-parametric insights to effectively overcome these limitations.

\subsection{Retrieval-Augmented Knowledge}
LLMs falter in deep-knowledge domains such as open-domain QA and reasoning, hindered by insufficiently encoded implicit knowledge.
The ICL mechanism avoids updating model parameters and improves performance with a few annotated examples~\cite{ICL/1, ICL/2, ICL/5,DBLP:conf/jist/0007HCGFP0Z22,DBLP:conf/icde/00070HCGYBZYSWY23}.
However, it struggles in fields like molecular science with unique linguistic structures.
Our work addresses the challenges LLMs face in specialized fields like molecular science, where data often features unique syntax beyond natural language texts~\cite{DBLP:journals/corr/abs-2305-14642,DBLP:conf/aaai/FangZYZD0Q0FC22,fang2023knowledge,DBLP:journals/corr/abs-2301-11259}. 
Acknowledging LLMs' proficiency with basic texts yet struggles with advanced expertise, our non-parametric DRAK strategy, encompassing template design, sample selection, and retrieval, adeptly navigates ICL limitations. 

\section{Conclusion}
% \paragraph{Conclusion.}
% This research introduces a non-parametric knowledge injection strategy that markedly advances the proficiency of LLMs within the biomedical realm. By seamlessly weaving domain-specific knowledge into the fabric of LLMs, our approach ensures a nuanced and context-aware analysis of molecular information, effectively bridging the gap in specialized syntax comprehension and application in biomedicine. 
% The application of two innovative prompting strategies has resulted in notable improvements in the performance of biomedical LLMs, achieving impressive results on six molecular tasks within the Mol-instructions dataset.
% These achievements confirm our method's effectiveness and underscore its potential to enhance LLMs in biomedicine. 
% Our work highlights the crucial role of targeted knowledge integration in understanding molecular structures, promising significant advances in drug discovery and related areas.
This study presents a non-parametric knowledge injection approach to bolster LLMs' performance in the biomolecular sector, integrating domain-specific insights for a better understanding and application of biomolecular science.
We analyzed the LLMs' understanding of different forms of molecular data and their sensitivity to context through two prompting techniques.
Demonstrated by significant enhancements in six molecular tasks within the Mol-Instructions dataset, our method highlights the benefits of tailored knowledge integration for LLMs in drug discovery and molecular studies. 

% Moreover, we observe that employing specialized tools enables a more accurate determination of molecular structures and characteristics. In our upcoming research endeavors, we will explore the integration of AI agents with molecular LLM assistants, aiming to significantly improve our understanding and accuracy in the realm of molecular interactions.

\section*{Limitations}
We introduce a new paradigm for molecular LLMs in question-answering, grounded in Domain-Specific Retrieval-Augmented Knowledge. However, ``the light of science, though bright, is not without its flaws.'' We acknowledge the necessity to discuss these potential limitations and constraints:

\paragraph{Dependence on External Databases and Tools:} DRAK's operation is intertwined with external databases (e.g., ChromaDB) and tools (e.g., RDKit, networkx). Any limitations, inaccuracies, or updates in these external resources directly impact DRAK's performance. Ensuring the continuous accuracy and relevance of these tools and databases is crucial for maintaining DRAK's effectiveness.

\paragraph{Adaptability to Rapidly Evolving Fields:} The biomolecular field is rapidly evolving, with new discoveries and insights emerging regularly. DRAK's current framework may not be agile enough to incorporate the latest research findings or changes in domain knowledge without significant updates or retraining, potentially leading to outdated or less accurate responses over time.

\paragraph{Generalization Across Domains:} DRAK has been primarily tested and validated in the context of biomolecules. Its ability to generalize across other knowledge-intensive domains, such as physics, engineering, or even within subdomains of biology, remains to be thoroughly explored. Each domain has its unique challenges, terminologies, and data formats, which may require adaptations to the DRAK framework for optimal performance.

\paragraph{Data Bias and Representation:} The model's performance is inherently tied to the quality and diversity of the training data. Biases present in the datasets—such as overrepresentation of certain molecule types or underrepresentation of rare molecular structures—can lead to skewed model outputs, potentially impacting the reliability of generated molecules or predictions.

\paragraph{Bioengineering Misuse:} The ability of DRAK to design new molecules and predict their functions could be exploited for the creation of harmful substances or bioweapons. Ensuring that such powerful capabilities are not misused is of utmost importance.

\section*{Mitigating Strategies}

\paragraph{Regulated Access and Usage Monitoring:} Implementing controlled access to DRAK for verified entities and monitoring usage patterns can help prevent misuse. Developing algorithms to detect and flag explorations into sensitive or potentially harmful domains is crucial.

\paragraph{Enhancing Data Quality and Diversity:} Efforts should be made to curate and diversify the datasets used for training DRAK. This includes addressing biases and ensuring a broad representation of molecular structures to improve the model's accuracy and generalizability.

\paragraph{Transparency in Model Development:} Adopting practices that increase the transparency of the AI's decision-making processes can enhance trust and reliability. This might involve developing more interpretable models or providing detailed documentation of the model's training data and algorithms.

In conclusion, while DRAK presents a significant advancement in applying LLMs to the domain of biomolecules, moving forward with an awareness of its limitations and potential ethical implications is essential. By implementing robust measures to mitigate risks and ensure ethical use, the benefits of such technologies can be realized while minimizing potential harms.

% \section*{Ethics Statement}

% \section*{Acknowledgements}

\bibliography{anthology,custom}

\begin{thebibliography}{49}
\expandafter\ifx\csname natexlab\endcsname\relax\def\natexlab#1{#1}\fi

\bibitem[{Banerjee and Lavie(2005)}]{METEOR}
Satanjeev Banerjee and Alon Lavie. 2005.
\newblock {METEOR:} an automatic metric for {MT} evaluation with improved correlation with human judgments.
\newblock In \emph{IEEvaluation@ACL}, pages 65--72. Association for Computational Linguistics.

\bibitem[{Blair{-}Stanek et~al.(2023)Blair{-}Stanek, Holzenberger, and Durme}]{survey/3}
Andrew Blair{-}Stanek, Nils Holzenberger, and Benjamin~Van Durme. 2023.
\newblock Can {GPT-3} perform statutory reasoning?
\newblock In \emph{{ICAIL}}, pages 22--31. {ACM}.

\bibitem[{Cao et~al.(2023)Cao, Liu, Lu, Yao, and Li}]{InstructMol}
He~Cao, Zijing Liu, Xingyu Lu, Yuan Yao, and Yu~Li. 2023.
\newblock Instructmol: Multi-modal integration for building a versatile and reliable molecular assistant in drug discovery.
\newblock \emph{CoRR}, abs/2311.16208.

\bibitem[{Chen et~al.(2021)Chen, Chen, Geng, Pan, Yuan, and Chen}]{DBLP:conf/semweb/0007CGPYC21}
Zhuo Chen, Jiaoyan Chen, Yuxia Geng, Jeff~Z. Pan, Zonggang Yuan, and Huajun Chen. 2021.
\newblock Zero-shot visual question answering using knowledge graph.
\newblock In \emph{{ISWC}}, volume 12922 of \emph{Lecture Notes in Computer Science}, pages 146--162. Springer.

\bibitem[{Chen et~al.(2022)Chen, Huang, Chen, Geng, Fang, Pan, Zhang, and Zhang}]{DBLP:conf/jist/0007HCGFP0Z22}
Zhuo Chen, Yufeng Huang, Jiaoyan Chen, Yuxia Geng, Yin Fang, Jeff~Z. Pan, Ningyu Zhang, and Wen Zhang. 2022.
\newblock Lako: Knowledge-driven visual question answering via late knowledge-to-text injection.
\newblock In \emph{{IJCKG}}, pages 20--29. {ACM}.

\bibitem[{Chen et~al.(2023)Chen, Zhang, Huang, Chen, Geng, Yu, Bi, Zhang, Yao, Song, Wu, Yang, Chen, Lian, Li, Cheng, and Chen}]{DBLP:conf/icde/00070HCGYBZYSWY23}
Zhuo Chen, Wen Zhang, Yufeng Huang, Mingyang Chen, Yuxia Geng, Hongtao Yu, Zhen Bi, Yichi Zhang, Zhen Yao, Wenting Song, Xinliang Wu, Yi~Yang, Mingyi Chen, Zhaoyang Lian, Yingying Li, Lei Cheng, and Huajun Chen. 2023.
\newblock Tele-knowledge pre-training for fault analysis.
\newblock In \emph{{ICDE}}, pages 3453--3466. {IEEE}.

\bibitem[{Chen et~al.(2024)Chen, Zhang, Fang, Geng, Guo, Chen, Li, Zhang, Chen, Zhu, Li, Liu, Pan, Zhang, and Chen}]{DBLP:journals/corr/abs-2402-05391}
Zhuo Chen, Yichi Zhang, Yin Fang, Yuxia Geng, Lingbing Guo, Xiang Chen, Qian Li, Wen Zhang, Jiaoyan Chen, Yushan Zhu, Jiaqi Li, Xiaoze Liu, Jeff~Z. Pan, Ningyu Zhang, and Huajun Chen. 2024.
\newblock Knowledge graphs meet multi-modal learning: {A} comprehensive survey.
\newblock \emph{CoRR}, abs/2402.05391.

\bibitem[{Chiang et~al.(2023)Chiang, Li, Lin, Sheng, Wu, Zhang, Zheng, Zhuang, Zhuang, Gonzalez, Stoica, and Xing}]{vicuna}
Wei-Lin Chiang, Zhuohan Li, Zi~Lin, Ying Sheng, Zhanghao Wu, Hao Zhang, Lianmin Zheng, Siyuan Zhuang, Yonghao Zhuang, Joseph~E. Gonzalez, Ion Stoica, and Eric~P. Xing. 2023.
\newblock \href {https://lmsys.org/blog/2023-03-30-vicuna/} {Vicuna: An open-source chatbot impressing gpt-4 with 90\%* chatgpt quality}.

\bibitem[{Dong et~al.(2023)Dong, Li, Dai, Zheng, Wu, Chang, Sun, Xu, Li, and Sui}]{ICL/1}
Qingxiu Dong, Lei Li, Damai Dai, Ce~Zheng, Zhiyong Wu, Baobao Chang, Xu~Sun, Jingjing Xu, Lei Li, and Zhifang Sui. 2023.
\newblock A survey for in-context learning.
\newblock \emph{CoRR}, abs/2301.00234.

\bibitem[{Durant et~al.(2002)Durant, Leland, Henry, and Nourse}]{Morgan}
Joseph~L. Durant, Burton~A. Leland, Douglas~R. Henry, and James~G. Nourse. 2002.
\newblock Reoptimization of {MDL} keys for use in drug discovery.
\newblock \emph{J. Chem. Inf. Comput. Sci.}, 42(5):1273--1280.

\bibitem[{Edwards et~al.(2022)Edwards, Lai, Ros, Honke, Cho, and Ji}]{MolT5}
Carl Edwards, Tuan~Manh Lai, Kevin Ros, Garrett Honke, Kyunghyun Cho, and Heng Ji. 2022.
\newblock Translation between molecules and natural language.
\newblock In \emph{{EMNLP}}, pages 375--413. Association for Computational Linguistics.

\bibitem[{Fang et~al.(2023{\natexlab{a}})Fang, Liang, Zhang, Liu, Huang, Chen, Fan, and Chen}]{Mol-Instructions}
Yin Fang, Xiaozhuan Liang, Ningyu Zhang, Kangwei Liu, Rui Huang, Zhuo Chen, Xiaohui Fan, and Huajun Chen. 2023{\natexlab{a}}.
\newblock Mol-instructions: {A} large-scale biomolecular instruction dataset for large language models.
\newblock \emph{CoRR}, abs/2306.08018.

\bibitem[{Fang et~al.(2023{\natexlab{b}})Fang, Zhang, Chen, Fan, and Chen}]{DBLP:journals/corr/abs-2301-11259}
Yin Fang, Ningyu Zhang, Zhuo Chen, Xiaohui Fan, and Huajun Chen. 2023{\natexlab{b}}.
\newblock Domain-agnostic molecular generation with self-feedback.
\newblock \emph{CoRR}, abs/2301.11259.

\bibitem[{Fang et~al.(2022{\natexlab{a}})Fang, Zhang, Chen, Fan, and Chen}]{DBLP:journals/corr/abs-2202-10587}
Yin Fang, Qiang Zhang, Zhuo Chen, Xiaohui Fan, and Huajun Chen. 2022{\natexlab{a}}.
\newblock Knowledge-informed molecular learning: {A} survey on paradigm transfer.
\newblock \emph{CoRR}, abs/2202.10587.

\bibitem[{Fang et~al.(2022{\natexlab{b}})Fang, Zhang, Yang, Zhuang, Deng, Zhang, Qin, Chen, Fan, and Chen}]{DBLP:conf/aaai/FangZYZD0Q0FC22}
Yin Fang, Qiang Zhang, Haihong Yang, Xiang Zhuang, Shumin Deng, Wen Zhang, Ming Qin, Zhuo Chen, Xiaohui Fan, and Huajun Chen. 2022{\natexlab{b}}.
\newblock Molecular contrastive learning with chemical element knowledge graph.
\newblock In \emph{{AAAI}}, pages 3968--3976. {AAAI} Press.

\bibitem[{Fang et~al.(2023{\natexlab{c}})Fang, Zhang, Zhang, Chen, Zhuang, Shao, Fan, and Chen}]{fang2023knowledge}
Yin Fang, Qiang Zhang, Ningyu Zhang, Zhuo Chen, Xiang Zhuang, Xin Shao, Xiaohui Fan, and Huajun Chen. 2023{\natexlab{c}}.
\newblock Knowledge graph-enhanced molecular contrastive learning with functional prompt.
\newblock \emph{Nature Machine Intelligence}, pages 1--12.

\bibitem[{Fergus et~al.(2023)Fergus, Botha, and Ostovar}]{bioLLM/3}
Suzanne Fergus, Michelle Botha, and Mehrnoosh Ostovar. 2023.
\newblock Evaluating academic answers generated using chatgpt.
\newblock \emph{Journal of Chemical Education}, 100(4):1672--1675.

\bibitem[{Guo et~al.(2023)Guo, Wang, Chen, Zhang, Sun, Lai, Zhang, and Chen}]{DBLP:journals/corr/abs-2305-14642}
Lingbing Guo, Weiqing Wang, Zhuo Chen, Ningyu Zhang, Zequn Sun, Yixuan Lai, Qiang Zhang, and Huajun Chen. 2023.
\newblock Newton-cotes graph neural networks: On the time evolution of dynamic systems.
\newblock \emph{CoRR}, abs/2305.14642.

\bibitem[{Guu et~al.(2020)Guu, Lee, Tung, Pasupat, and Chang}]{RAK/2}
Kelvin Guu, Kenton Lee, Zora Tung, Panupong Pasupat, and Ming{-}Wei Chang. 2020.
\newblock Retrieval augmented language model pre-training.
\newblock In \emph{{ICML}}, volume 119 of \emph{Proceedings of Machine Learning Research}, pages 3929--3938. {PMLR}.

\bibitem[{Hu et~al.(2022)Hu, Shen, Wallis, Allen{-}Zhu, Li, Wang, Wang, and Chen}]{lora}
Edward~J. Hu, Yelong Shen, Phillip Wallis, Zeyuan Allen{-}Zhu, Yuanzhi Li, Shean Wang, Lu~Wang, and Weizhu Chen. 2022.
\newblock Lora: Low-rank adaptation of large language models.
\newblock In \emph{{ICLR}}. OpenReview.net.

\bibitem[{Huang et~al.(2023)Huang, Yu, Ma, Zhong, Feng, Wang, Chen, Peng, Feng, Qin, and Liu}]{hallucination/2}
Lei Huang, Weijiang Yu, Weitao Ma, Weihong Zhong, Zhangyin Feng, Haotian Wang, Qianglong Chen, Weihua Peng, Xiaocheng Feng, Bing Qin, and Ting Liu. 2023.
\newblock A survey on hallucination in large language models: Principles, taxonomy, challenges, and open questions.
\newblock \emph{CoRR}, abs/2311.05232.

\bibitem[{Krenn et~al.(2020)Krenn, H{\"{a}}se, Nigam, Friederich, and Aspuru{-}Guzik}]{SELFIES}
Mario Krenn, Florian H{\"{a}}se, AkshatKumar Nigam, Pascal Friederich, and Al{\'{a}}n Aspuru{-}Guzik. 2020.
\newblock Self-referencing embedded strings {(SELFIES):} {A} 100{\%} robust molecular string representation.
\newblock \emph{Mach. Learn. Sci. Technol.}, 1(4):45024.

\bibitem[{Landrum et~al.(2013)}]{RDKit}
Greg Landrum et~al. 2013.
\newblock Rdkit: A software suite for cheminformatics, computational chemistry, and predictive modeling.
\newblock \emph{Greg Landrum}, 8:31.

\bibitem[{Lewis et~al.(2020)Lewis, Perez, Piktus, Petroni, Karpukhin, Goyal, K{\"{u}}ttler, Lewis, Yih, Rockt{\"{a}}schel, Riedel, and Kiela}]{RAK/1}
Patrick S.~H. Lewis, Ethan Perez, Aleksandra Piktus, Fabio Petroni, Vladimir Karpukhin, Naman Goyal, Heinrich K{\"{u}}ttler, Mike Lewis, Wen{-}tau Yih, Tim Rockt{\"{a}}schel, Sebastian Riedel, and Douwe Kiela. 2020.
\newblock Retrieval-augmented generation for knowledge-intensive {NLP} tasks.
\newblock In \emph{NeurIPS}.

\bibitem[{Li and Liu(2007)}]{Levenshtein}
Yujian Li and Bi~Liu. 2007.
\newblock A normalized levenshtein distance metric.
\newblock \emph{{IEEE} Trans. Pattern Anal. Mach. Intell.}, 29(6):1091--1095.

\bibitem[{Liang et~al.(2023)Liang, Zhang, Zhang, and Xie}]{bioLLM/2}
Youwei Liang, Ruiyi Zhang, Li~Zhang, and Pengtao Xie. 2023.
\newblock Drugchat: Towards enabling chatgpt-like capabilities on drug molecule graphs.
\newblock \emph{CoRR}, abs/2309.03907.

\bibitem[{Lin(2004)}]{ROUGE}
Chin-Yew Lin. 2004.
\newblock Rouge: A package for automatic evaluation of summaries.
\newblock In \emph{Text summarization branches out}, pages 74--81.

\bibitem[{Lin et~al.(2023)Lin, Tang, Tang, Yang, Dang, and Han}]{ICL/5}
Ji~Lin, Jiaming Tang, Haotian Tang, Shang Yang, Xingyu Dang, and Song Han. 2023.
\newblock {AWQ:} activation-aware weight quantization for {LLM} compression and acceleration.
\newblock \emph{CoRR}, abs/2306.00978.

\bibitem[{Liu et~al.(2022)Liu, Shen, Zhang, Dolan, Carin, and Chen}]{liu2021makes}
Jiachang Liu, Dinghan Shen, Yizhe Zhang, Bill Dolan, Lawrence Carin, and Weizhu Chen. 2022.
\newblock What makes good in-context examples for gpt-3?
\newblock In \emph{DeeLIO@ACL}, pages 100--114. Association for Computational Linguistics.

\bibitem[{Liu et~al.(2023)Liu, Nie, Wang, Lu, Qiao, Liu, Tang, Xiao, and Anandkumar}]{Multi-modal/1}
Shengchao Liu, Weili Nie, Chengpeng Wang, Jiarui Lu, Zhuoran Qiao, Ling Liu, Jian Tang, Chaowei Xiao, and Animashree Anandkumar. 2023.
\newblock Multi-modal molecule structure-text model for text-based retrieval and editing.
\newblock \emph{Nat. Mac. Intell.}, 5(12):1447--1457.

\bibitem[{Papineni et~al.(2002)Papineni, Roukos, Ward, and Zhu}]{BLEU}
Kishore Papineni, Salim Roukos, Todd Ward, and Wei{-}Jing Zhu. 2002.
\newblock Bleu: a method for automatic evaluation of machine translation.
\newblock In \emph{{ACL}}, pages 311--318. {ACL}.

\bibitem[{Qian et~al.(2023)Qian, Tang, Yang, Liang, and Liu}]{bioLLM/1}
Chen Qian, Huayi Tang, Zhirui Yang, Hong Liang, and Yong Liu. 2023.
\newblock Can large language models empower molecular property prediction?
\newblock \emph{CoRR}, abs/2307.07443.

\bibitem[{Schneider et~al.(2015)Schneider, Sayle, and Landrum}]{MACCS}
Nadine Schneider, Roger~A. Sayle, and Gregory~A. Landrum. 2015.
\newblock Get your atoms in order - an open-source implementation of a novel and robust molecular canonicalization algorithm.
\newblock \emph{J. Chem. Inf. Model.}, 55(10):2111--2120.

\bibitem[{Su et~al.(2022)Su, Du, Yang, Zhou, Li, Rao, Sun, Lu, and Wen}]{Multi-modal/2}
Bing Su, Dazhao Du, Zhao Yang, Yujie Zhou, Jiangmeng Li, Anyi Rao, Hao Sun, Zhiwu Lu, and Ji{-}Rong Wen. 2022.
\newblock A molecular multimodal foundation model associating molecule graphs with natural language.
\newblock \emph{CoRR}, abs/2209.05481.

\bibitem[{Tanimoto(1958)}]{tanimoto}
Taffee~T Tanimoto. 1958.
\newblock Elementary mathematical theory of classification and prediction.

\bibitem[{Tloen(2023)}]{alpaca-lora}
Tloen. 2023.
\newblock Alpaca-lora.
\newblock \url{https://github.com/tloen/alpaca-lora}.

\bibitem[{Touvron et~al.(2023)Touvron, Lavril, Izacard, Martinet, Lachaux, Lacroix, Rozi{\`{e}}re, Goyal, Hambro, Azhar, Rodriguez, Joulin, Grave, and Lample}]{llama}
Hugo Touvron, Thibaut Lavril, Gautier Izacard, Xavier Martinet, Marie{-}Anne Lachaux, Timoth{\'{e}}e Lacroix, Baptiste Rozi{\`{e}}re, Naman Goyal, Eric Hambro, Faisal Azhar, Aur{\'{e}}lien Rodriguez, Armand Joulin, Edouard Grave, and Guillaume Lample. 2023.
\newblock Llama: Open and efficient foundation language models.
\newblock \emph{CoRR}, abs/2302.13971.

\bibitem[{Wan et~al.(2023)Wan, Cheng, Mao, Liu, Song, Li, and Kurohashi}]{wan2023gpt}
Zhen Wan, Fei Cheng, Zhuoyuan Mao, Qianying Liu, Haiyue Song, Jiwei Li, and Sadao Kurohashi. 2023.
\newblock {GPT-RE:} in-context learning for relation extraction using large language models.
\newblock In \emph{{EMNLP}}, pages 3534--3547. Association for Computational Linguistics.

\bibitem[{Wei et~al.(2022)Wei, Wang, Schuurmans, Bosma, Ichter, Xia, Chi, Le, and Zhou}]{CoT}
Jason Wei, Xuezhi Wang, Dale Schuurmans, Maarten Bosma, Brian Ichter, Fei Xia, Ed~H. Chi, Quoc~V. Le, and Denny Zhou. 2022.
\newblock Chain-of-thought prompting elicits reasoning in large language models.
\newblock In \emph{NeurIPS}.

\bibitem[{Wei et~al.(2023)Wei, Wei, Tay, Tran, Webson, Lu, Chen, Liu, Huang, Zhou, and Ma}]{ICL/2}
Jerry~W. Wei, Jason Wei, Yi~Tay, Dustin Tran, Albert Webson, Yifeng Lu, Xinyun Chen, Hanxiao Liu, Da~Huang, Denny Zhou, and Tengyu Ma. 2023.
\newblock Larger language models do in-context learning differently.
\newblock \emph{CoRR}, abs/2303.03846.

\bibitem[{Weininger(1988)}]{SMILES}
David Weininger. 1988.
\newblock Smiles, a chemical language and information system. 1. introduction to methodology and encoding rules.
\newblock \emph{J. Chem. Inf. Comput. Sci.}, 28(1):31--36.

\bibitem[{White(2020)}]{pubmed}
Jacob White. 2020.
\newblock Pubmed 2.0.
\newblock \emph{Medical reference services quarterly}, 39(4):382--387.

\bibitem[{Xu et~al.(2023)Xu, Guo, Duan, and McAuley}]{Baize}
Canwen Xu, Daya Guo, Nan Duan, and Julian~J. McAuley. 2023.
\newblock Baize: An open-source chat model with parameter-efficient tuning on self-chat data.
\newblock In \emph{{EMNLP}}, pages 6268--6278. Association for Computational Linguistics.

\bibitem[{Zeng et~al.(2023)Zeng, Liu, Du, Wang, Lai, Ding, Yang, Xu, Zheng, Xia, Tam, Ma, Xue, Zhai, Chen, Liu, Zhang, Dong, and Tang}]{Glm}
Aohan Zeng, Xiao Liu, Zhengxiao Du, Zihan Wang, Hanyu Lai, Ming Ding, Zhuoyi Yang, Yifan Xu, Wendi Zheng, Xiao Xia, Weng~Lam Tam, Zixuan Ma, Yufei Xue, Jidong Zhai, Wenguang Chen, Zhiyuan Liu, Peng Zhang, Yuxiao Dong, and Jie Tang. 2023.
\newblock {GLM-130B:} an open bilingual pre-trained model.
\newblock In \emph{{ICLR}}. OpenReview.net.

\bibitem[{Zhang et~al.(2023{\natexlab{a}})Zhang, Dong, Li, Zhang, Sun, Wang, Li, Hu, Zhang, Wu, and Wang}]{survey/tuneing/1}
Shengyu Zhang, Linfeng Dong, Xiaoya Li, Sen Zhang, Xiaofei Sun, Shuhe Wang, Jiwei Li, Runyi Hu, Tianwei Zhang, Fei Wu, and Guoyin Wang. 2023{\natexlab{a}}.
\newblock Instruction tuning for large language models: {A} survey.
\newblock \emph{CoRR}, abs/2308.10792.

\bibitem[{Zhang et~al.(2023{\natexlab{b}})Zhang, Dong, Li, Zhang, Sun, Wang, Li, Hu, Zhang, Wu, and Wang}]{survey/instruction}
Shengyu Zhang, Linfeng Dong, Xiaoya Li, Sen Zhang, Xiaofei Sun, Shuhe Wang, Jiwei Li, Runyi Hu, Tianwei Zhang, Fei Wu, and Guoyin Wang. 2023{\natexlab{b}}.
\newblock Instruction tuning for large language models: {A} survey.
\newblock \emph{CoRR}, abs/2308.10792.

\bibitem[{Zhang et~al.(2023{\natexlab{c}})Zhang, Chen, Fang, Cheng, Lu, Li, Zhang, and Chen}]{zhang2023knowledgeable}
Yichi Zhang, Zhuo Chen, Yin Fang, Lei Cheng, Yanxi Lu, Fangming Li, Wen Zhang, and Huajun Chen. 2023{\natexlab{c}}.
\newblock Knowledgeable preference alignment for llms in domain-specific question answering.
\newblock \emph{CoRR}, abs/2311.06503.

\bibitem[{Zhang et~al.(2023{\natexlab{d}})Zhang, Chen, Zhang, and Chen}]{DBLP:journals/corr/abs-2310-06671}
Yichi Zhang, Zhuo Chen, Wen Zhang, and Huajun Chen. 2023{\natexlab{d}}.
\newblock Making large language models perform better in knowledge graph completion.
\newblock \emph{CoRR}, abs/2310.06671.

\bibitem[{Zhang et~al.(2023{\natexlab{e}})Zhang, Li, Cui, Cai, Liu, Fu, Huang, Zhao, Zhang, Chen, Wang, Luu, Bi, Shi, and Shi}]{hallucination/1}
Yue Zhang, Yafu Li, Leyang Cui, Deng Cai, Lemao Liu, Tingchen Fu, Xinting Huang, Enbo Zhao, Yu~Zhang, Yulong Chen, Longyue Wang, Anh~Tuan Luu, Wei Bi, Freda Shi, and Shuming Shi. 2023{\natexlab{e}}.
\newblock Siren's song in the {AI} ocean: {A} survey on hallucination in large language models.
\newblock \emph{CoRR}, abs/2309.01219.

\end{thebibliography}
\bibliographystyle{acl_natbib}
\appendix

% \begin{table*}[t!]
%     \centering
%     \renewcommand\arraystretch{1}
%     \scalebox{1.}{
%     \begin{tabular}{ccc}
%     \toprule
%     \textbf{Name} &  \textbf{Llama-2-7b\&13b-chat} & \textbf{Llama-2-70b-chat} \\
%     \Xhline{1px}
%     lora\_r & 8 & 8 \\
%     lora\_alpha & 16 & 16 \\
%     lora\_dropout & 0.05 & 0.05 \\
%     lora\_target\_modules & q\_proj, v\_proj & q\_proj, v\_proj \\
%     model\_max\_length & 4096 & 4096 \\
%     per\_device\_batch\_size & 2 & 2 \\
%     gradient\_accumulation\_steps & 1 & 1 \\
%     warmup\_ratio & 0.03 & 0.03 \\
%     epochs & 5 & 3 \\
%     batch size & 4 & 1 \\
%     learning rate & 1e-4 & 1e-4 \\
%     \bottomrule
%     \end{tabular}
%     }
%     \caption{Detailed hyper-parameters we use for training.}
%     \label{tab:hp}
% \end{table*}

\end{document}